# A 'normal' research trap limits scientific breakthroughs and disruptive innovation in the European Union

Alonso Rodríguez-Navarro

*Departamento de Biotecnología-Biología Vegetal, Universidad Politécnica de Madrid, Avenida Puerta de Hierro 2, 28040, Madrid, Spain*

*E-mail address:* *alonso.rodriguez@upm.es* ORCID 0000-0003-0462-223X

**Abstract**

Europe is experiencing prolonged slow growth, resulting in part from a shortage of disruptive innovations that has locked its industry into a "middle-technology trap." Although greater investment and deeper integration among EU member states are widely proposed as solutions, this study argues that these measures alone will be insufficient if breakthrough research is not improved. In research systems, breakthrough discoveries depend on both size and efficiency. The EU research system is constrained by a dual problem: it is substantially smaller than China's and significantly less efficient than that of the United States. As a result, it generates too few breakthrough discoveries to sustain competitiveness in high-technology sectors. Compared with the United Kingdom, the persistent underperformance of the EU's largest research systems—Germany, France, Italy, and Spain—has long been evident, yet meaningful reforms have not been implemented. Meanwhile, China's capacity for breakthrough research continues to expand rapidly. Despite sustained academic criticism, the European Commission has largely based its research policy on overly optimistic assessments of research performance that satisfy policymakers and researchers but fail to address structural weaknesses. Achieving a more balanced global technological landscape and strengthening Europe's competitiveness will require not only greater integration among member states but also a substantial improvement in research efficiency. Without such gains, the EU is unlikely to attract the private investment needed for high-technology industries, while additional public spending risks diverting resources from other sectors without addressing the underlying causes of Europe's innovation deficit.

## 1. Introduction

Disruptive innovations, which stem from breakthrough research, drive technological change and economic development (Feder, 2018). Breakthrough research accounts for only an insignificant proportion of all research output, the vast majority of which constitutes 'normal' research (Kuhn, 1970). This small proportion can be assessed through citation analysis, but if certain methodological rules are overlooked, the resulting assessments can be misleading. This is the case for Japan, a country with numerous Nobel Prize winners that nevertheless appears in most international research rankings as a scientifically developing country (Pendlebury, 2020). Such misleading assessments of research (Rodríguez-Navarro, 2025c; Rodríguez-Navarro & Brito, 2024a) can lead to misguided research policies and, consequently, hinder disruptive innovation.

### *1.1. Research and innovation in the EU*

The innovation deficit of the EU in advanced technologies has long been recognized, and it is widely acknowledged that this deficit constrains the EU economy. In 1995, the European Commission (EC) stated that "over the last fifteen years its technological and commercial performance in high-technology sectors such as electronics and information technologies has deteriorated" (European-Commission, 1995, p. 14). In 2018, it further acknowledged that "Disruptive and breakthrough innovations are still too rare in Europe" (European Commission, 2018, p. 11). This concern about high-technology sectors has persisted ever since. In 2024, Fuest et al. (Fuest et al., 2024) described the EU as being caught in a "middle-technology trap." Also in 2024, an extensive report by the former President of the European Central Bank, Mario Draghi, reached a similar conclusion: "EU companies are specialised in mature technologies where the potential for breakthroughs is limited" (Draghi, 2024a, p. 2).

The explanation most frequently offered for the EU's innovation gap is insufficient investment. Accordingly, the Draghi Report proposes substantial investment to close the innovation gap (Draghi, 2024a, p. 59). Similarly, the International Monetary Fund describes the innovation gap primarily in terms of investment: "The EU could nurture innovative startups by accelerating the development of its venture capital (VC) ecosystem" (Arnold et al., 2024, p. 0). Other explanations have also been proposed, including a "deep-rooted aversion to risk" (European Commission, 2018, p. 11) and "insufficient incentives for researchers to commercialize their work" (European Commission, 2025c, p. 10). However, the explanation most frequently emphasized is "a lack of transfer of new technologies from the research base," although "the EU is a global leader in many areas of science and technology" (European Commission, 2018, pp. 11 and 5, respectively). This view is equivalent to the statement that "Europe has a longstanding tradition of academic excellence, but we sometimes struggle to turn research into commercial success" (European Patent Office, 2024).

These widely held beliefs stem from the EC's longstanding characterization of EU research as excellent. Since the *Green Paper on Innovation* in 1995, the EC has argued that "the situation of the European Union in terms of innovation appears to be unsatisfactory, despite some first-rate scientific achievements" (European Commission, 1995, p. 5), when the expression *European Paradox* was coined. More recently, the EC has popularized statements such as "Europe is a global scientific powerhouse," "the contrast is striking between Europe's comparative advantage in producing knowledge and its comparative disadvantage in turning that knowledge into innovation and growth," and "Europe is a world leader in science" (Rodríguez-Navarro & Brito, 2020).

From a different perspective, Gros et al. (2025) argue that "the EU investment gap is really an R&D gap" (p. 667), although they do not define this research gap precisely. They note that "The EU corporate sector as a whole invests as much as the US one. The real problem is that the EU ecosystem does not provide the right incentives to invest in the high-risk, high-reward projects that are typically pursued in high-tech industries" (p. 677). Similarly, Draghi's report (2024a, 2024b) describes the EU's shortcomings in terms of a deficit in research excellence: "There are not enough academic institutions achieving top levels of excellence and the pipeline from innovation into commercialisation is weak" (Draghi, 2024a, p. 24); "It is also essential to establish and consolidate European academic institutions at the forefront of global research" (p. 29); and "Excellence in research and innovation is fundamental to the EU's competitiveness in a global economy where technological leaders have the ability to capture huge market shares" (p. 245). However, this argument is qualified elsewhere in the report: "Europe has a strong position in fundamental research" (p. 25), and "the robust scientific position is not fully reflected in its presence in innovative markets" (p. 231).

The existence of an "R&D gap" appears inconsistent with the EU's large volume of scientific publications, which exceeds that of the USA. This apparent contradiction requires defining the gap in terms of breakthrough research rather than total research output (Rodríguez-Navarro, 2025a; Rodríguez-Navarro & Brito, 2018). Figure 1 visually summarizes the deficit reported in these studies by comparing the citation distributions of EU and US domestic publications on graphene during the five-year period 2017–2021. The EU published approximately 60% more papers than the USA—8,964 compared with 5,685—but most of these publications are concentrated in the low-citation portion of the distribution. In contrast, the dominance of the USA in the top 10% of the citation distribution is evident, while its dominance in the top 1% is overwhelming.

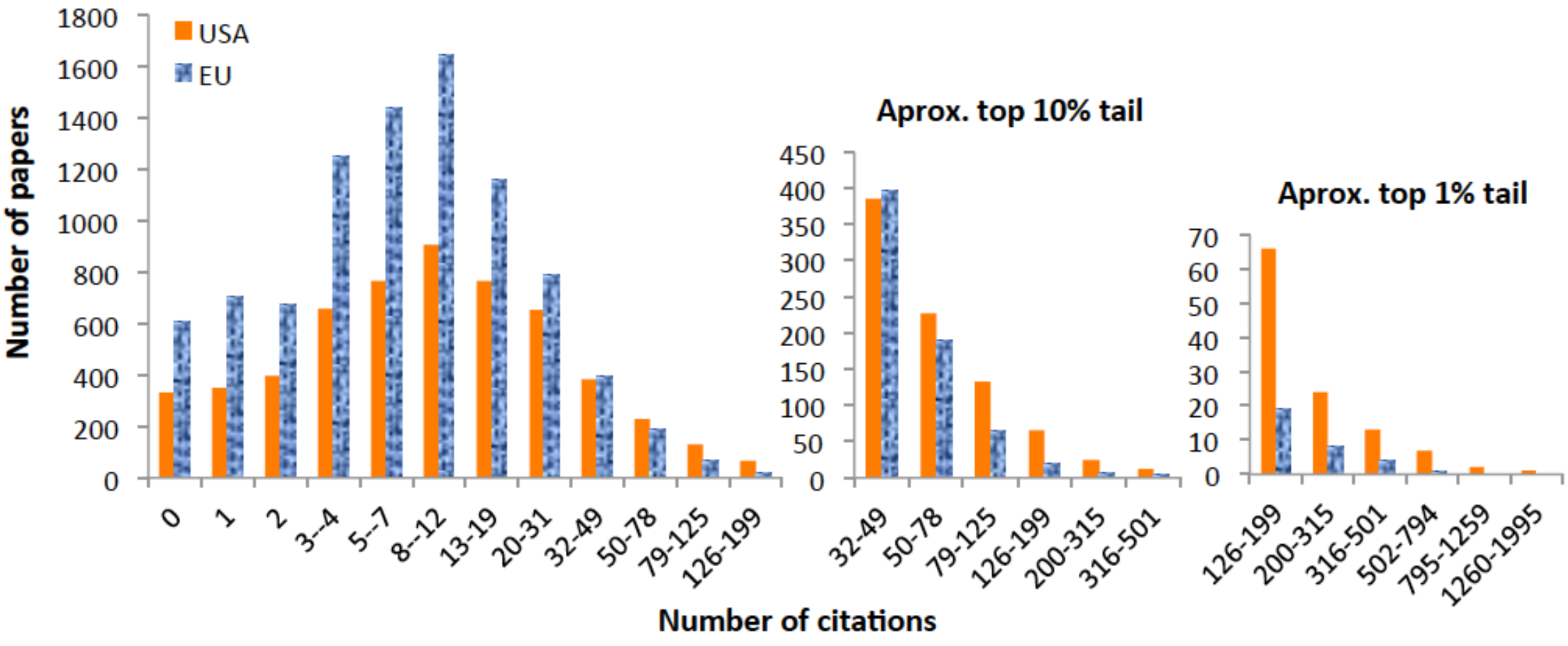

Fig. 1. Citation distribution with logarithmic binning of publications on graphene from the USA and the EU: distributions of all papers and of the upper 10% and 1% tails, approximately. Publication window, 2017–2021; citation window, 2023–2025.

*1.2. Wrong diagnoses and misguided policies*

In 2006, Dosi et al. (2006, p. 1461) used the title of this section to describe the erroneous assumptions underlying the *European Paradox*. Since then, the EC has published a large number of reports without taking into account the research deficiencies identified in numerous academic publications (Argyropoulou et al., 2019). This disregard for the academic literature appears to stem from a form of scientific overconfidence that has led the EC to consistently proclaim an illusory excellence of EU research. For example, the report of a High-Level Group (European Commission, 2017) does not even cite the seminal and highly cited paper by Dosi et al. (2006).

More recently, in 2024 and 2025, the EC published three reports on Horizon Europe (European Commission, 2024a, 2025b, 2025d) without mentioning the shortcomings of EU research identified in the academic literature. Although these reports cite the Draghi Report (2024a), they do not cite the report by Fuest et al. (2024), which is critical of the Horizon Europe program.

Notably, one of these EC reports reiterates the *European Paradox* by referring to "the EU's longstanding weakness in bringing its excellent research to the market" (European Commission, 2025a, p. 8). Another report (European Commission, 2024a) assumes that Europe's global importance in research, innovation, and technological development has declined relative to 2017, when the EC declared that "Europe is a global scientific powerhouse" (p. 14). This assumption again ignores the academic literature published before 2017 demonstrating the flaws in the "scientific powerhouse" narrative (Rodríguez-Navarro & Brito, 2020).

### *1.3. Types of research*

In most reports on innovation, research is treated as a homogeneous activity, which hinders the analysis of technological research.

The *Frascati Manual* (OECD, 2015) distinguishes three types of R&D: basic research, applied research, and experimental development. For citation-based studies, however, it is convenient to combine the first two categories into a single type encompassing all research aimed at pushing the boundaries of knowledge. This combination is useful for two reasons. First, the boundary between science and technology is often difficult to define—for example, in lithium battery research (Feng et al., 2025) basic and applied research cannot be readily distinguished. Second, citation distributions reveal only two distinct types of publications.

It is widely accepted that citations to scientific papers follow a lognormal distribution (Golosovsky, 2021; Viiu, 2018), including a certain proportion of uncited papers. Although several mathematical explanations have been proposed for the substantial proportion of uncited papers (Golosovsky & Larivière, 2021; Thelwall, 2016), other studies have shown that uncited papers constitute only part of a larger group of publications that deviate from the lognormal distribution in the lower tail. The number of papers included in this deviation varies across disciplines. It is greater in highly technological fields and in countries with strong technological and patenting activity, such as Japan and South Korea (Rodríguez-Navarro, 2025c). On the contrary, journals that publish only research aimed at advancing the frontiers of knowledge exhibit the smallest deviations from the lognormal distribution (Rodríguez-Navarro, 2024).

These findings demonstrate the existence of two populations of publications that can be distinguished by their citation distributions: one following a lognormal distribution and another following a distribution that decreases monotonically from the number of uncited papers. Figure 2 presents two examples of the citation distributions of worldwide publications in two disciplines—immunology and semiconductors—with, respectively, small and large deviations from the lognormal distribution in the lower tail (reproduced from Rodríguez-Navarro, 2025c). The pronounced lower-tail deviation observed in semiconductors appears to correspond to research that primarily supports incremental innovation.

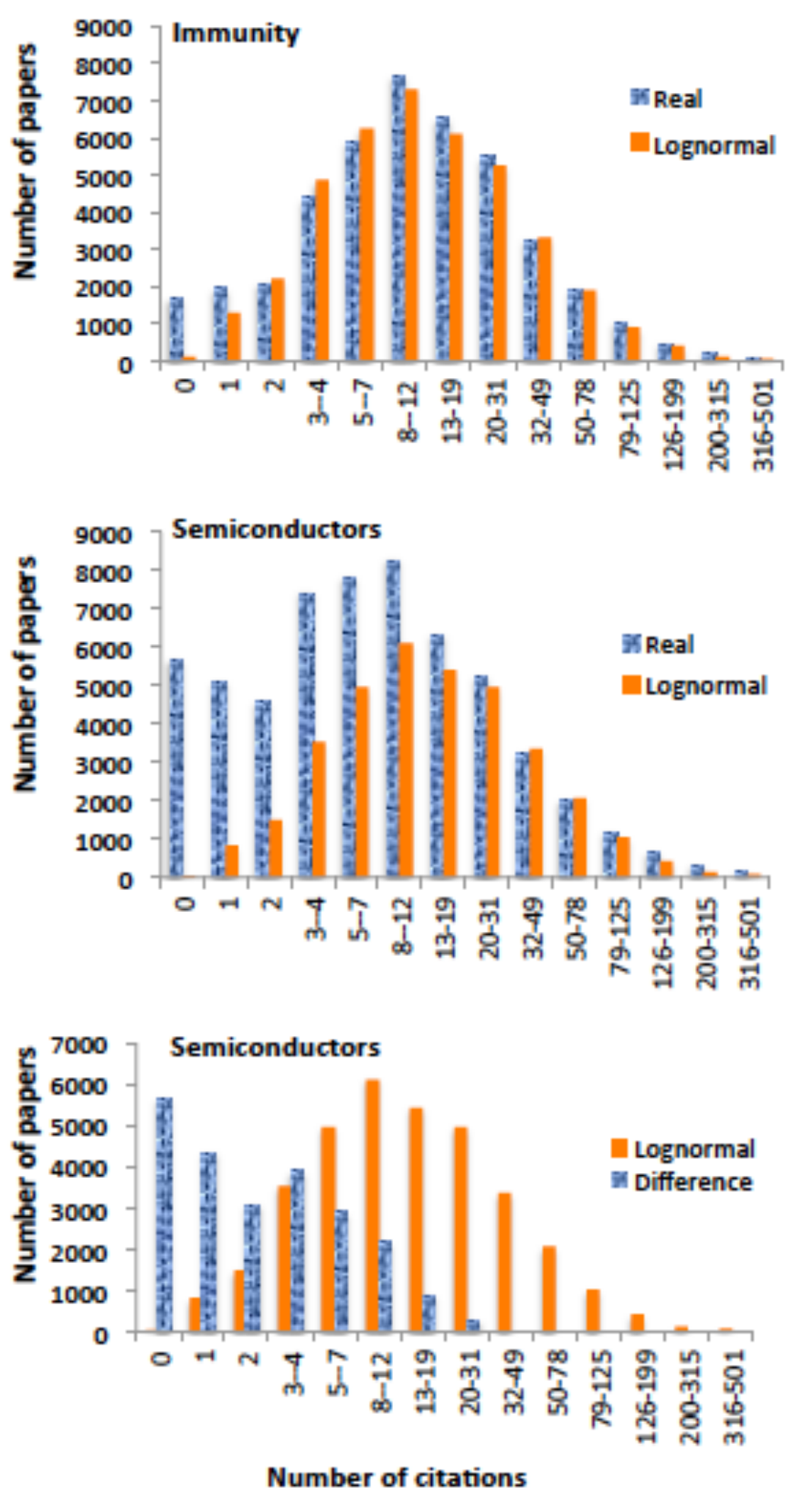

**Fig. 2.** Citation distribution with logarithmic binning of global publications in immunity and semiconductors. The lognormal distributions have been fitted to guide the eye. The bottom plot shows the lognormal distribution and the difference between the real data and the lognormal distribution for semiconductor publications. The two upper plots are reproduced from [Rodríguez-Navarro, 2025 #2153]. Publication window, 2014–2017; citation window, 2019–2022.

Failing to recognize the existence of these two types of research and their corresponding citation distributions can lead to misleading conclusions. For example, the misleading evaluations of Japan mentioned above (Section 1) result from ignoring the existence of publications whose citation distributions do not belong to the lognormal distribution (Rodríguez-Navarro, 2025c).

*1.4. Disruptive innovations arise from groundbreaking research*

Numerous studies have demonstrated the relationship between highly cited papers and the value of patents (Ahmadpoor & Jones, 2017; Breschi et al., 2006; Hicks et al., 2000; Narin et al., 1997; Narin & Olivastro, 1998; Poege et al., 2019). Papers that rank among the top 0.01% by citations in scientific journals have a high probability of being cited by patents and of generating patents with high commercial value (Poege et al., 2019). This implies that the number of these landmark papers (Bornmann et al., 2018) can be used to evaluate a country's capacity to generate disruptive innovations. In other words, countries with poor performance in the top 0.01% of cited publications are unlikely to succeed in producing disruptive innovations (Rodríguez-Navarro, 2025a; Rodríguez-Navarro & Brito, 2018). As shown in Fig. 1, the small difference between EU and US research performance at the top 10% level increases substantially at the top 1% level and becomes much greater at the top 0.01% level (Rodríguez-Navarro & Brito, 2018).

The link between breakthrough research and disruptive innovation has an important peculiarity because, in many cases, only the first to achieve the scientific breakthrough captures the full reward of the resulting innovation. The COVID-19 pandemic illustrates this phenomenon. During 2019–2020, most developed countries produced a large amount of 'normal' research in the search for a vaccine, but only a few succeeded in commercializing one. This success was based on mRNA vaccines (Vebeke et al., 2019) and lipid nanoparticles (Gregoriadis, 2021), and researchers in the countries with the strongest breakthrough research in these technologies ultimately prevailed.

When considering research success, one intuitively thinks of efficiency—that is, the research quality that results in a larger proportion of US than EU papers in the top 1% tail shown in Fig. 1. However, success also depends on size. If the EU had published six times as many papers as the USA—as is the case for China—the EU would have had more papers in the top 1% tail than the USA despite its lower efficiency. This effect of size can be understood also in probabilistic terms. Suppose that a bag contains 50,000 publications, of which only five are breakthrough papers. Each publication produced by a country is equivalent to randomly drawing a paper from the bag. For equal efficiency, the greater the number of draws, the higher the probability of selecting a breakthrough paper.

Assuming that breakthrough papers are produced, disruptive innovations depend on another capability: transferring scientific knowledge into technological applications.

Therefore, only when breakthrough research is produced efficiently can weak disruptive innovation be attributed to a limited capacity to transfer research into the commercial realm. The *European Paradox* assumes a flaw in the transfer by imagining that the research is excellent.

Not surprisingly, countries with the greatest capacity to produce breakthrough research also provide the strongest environment for commercializing scientific discoveries. Venture capital seeking disruptive innovations is unlikely to flow to countries with weak performance in breakthrough research.

### *1.5. Functions of 'normal' research*

The "normal research trap" of the EU described in this study is analogous to the "middle technology trap" (Fuest et al., 2024) and is based on Kuhn's definition of 'normal' research (Kuhn, 1970), which does not imply low-quality research.

Unfortunately, in citation-based research assessments, the actual number of 'normal' papers is unknown because the two populations of papers shown in Fig. 2 overlap considerably and the statistical distinction between the two citation distributions has not been established. In these circumstances, 'normal' research has to be assessed with indicators, and a reasonable indicator is the number of papers in the top 30% by citations (Rodríguez-Navarro, 2024). However, because this metric is not calculated in most country rankings, the proportion of papers in the top 10% by citations is a reasonable alternative to use together with the top 0.01% as an indicator of breakthrough research. The number of papers in this latter percentile is too small to be counted directly but can be statistically estimated from the citation distribution of the top 10% most cited papers (Section 3).

A country is anchored in "normal research trap" when its inability to produce breakthrough research persists over time. A country may never publish a breakthrough paper because it is either inefficient or too small, producing too few papers to have a reasonable chance of generating a breakthrough before larger countries do so (Section 1.3).

This reasoning raises the question of the role of 'normal' research in a country where the probability of producing breakthrough papers is very low. Certainly, normal research provides the foundation from which breakthrough papers emerge, but a country that produces only 'normal' papers may ultimately be investing in breakthroughs, disruptive innovations, and new products that are realized by other countries. This question applies specifically to research aimed at pushing the frontiers of knowledge and excludes research that supports incremental innovation, which requires separate consideration.

This question regarding the role of 'normal' research does not have a single answer because it depends on the type of product. For example, in the search for a vaccine during a new pandemic (Section 1.3), success depends on developing the vaccine. The normal research conducted in countries that failed to develop a vaccine ultimately provided little direct benefit to those countries. Other cases differ. For example, 'normal' research on new cancer treatments may include both the search for entirely new therapies and efforts to improve existing treatments, which may yield valuable outcomes even in the absence of breakthroughs.

## 2. Aims of this study

Currently, the EU is anchored in a "normal research trap," characterized by many sound papers and very few groundbreaking ones. This study aims to describe EU research at the breakthrough level in comparison with that of the USA, which is of similar size but much more competitive (Section 1.1), and with that of China, which has a larger research system. The question is whether the EU could compete with China in emerging technologies in the near future or should instead assume Chinese technological dominance and organize its economy and research accordingly.

The first part of the study focuses on universities, which can reasonably be assumed to represent the research level of a country. Indeed, most Nobel laureates conducted their prize-winning work at universities (Schlagberger et al., 2016). The second part examines specific technologies.

## 3. Materials and Methods

To obtain an overall view of a country's research performance, the *Leiden Ranking* is an excellent source of data, providing a comprehensive perspective based on a large number of microfields (https://www.leidenranking.com/information/fields). All data in this study were obtained from the *Leiden Ranking* 2024, using fractional counting (https://www.leidenranking.com/ranking/2024/; accessed in June 2025). The data correspond to a four-year publication window and are recorded in the second year after the end of the publication period (e.g., the 2024 ranking corresponds to the 2019–2022 publication period).

Hereinafter, I use the nomenclature of the *Leiden Ranking*: P is the total number of papers and $P_{top\ x\%}$ is the number of publications in the top x% most cited papers. A country's $P_{top\ x\%}$ is the sum of the $P_{top\ x\%}$ values of its universities.

The *Leiden Ranking* 2024 includes only universities that produced at least 800 publications indexed in the Web of Science. I focused on the field of *Physical Sciences and Engineering*. This field offers two advantages. First, it includes technologies of the greatest economic importance. Second, the dominance of US research in EU–US collaborative papers, which may lead to misleading indicators, has less effect in this field than in others (Rodríguez-Navarro, 2025b).

To investigate the research performance of a country with respect to research aimed at pushing the boundaries of knowledge, analysis of the top 10% tail of the most cited papers (Fig. 1) is an excellent tool. In research systems that follow an ideal model all $P_{top\ x\%}$/Ptop $_{(10*x)\%}$ ratios are equal (Rodríguez-Navarro, 2025c). However, several deviations from this ideal model occur; among these deviations, the existence of two populations of papers, as shown in Fig. 2, is the most frequent cause.

Three parameters can be derived from the top 10% tail: (i) $P_{top\ 10\%}$, an indicator of 'normal' research (Section 1.5); (ii) $P_{top\ 1\%}/P_{top\ 10\%}$, a size-independent indicator of the efficiency of a research system in producing highly cited papers; (iii) $P_{top\ 0.01\%}$, which estimates the expected number of breakthrough papers in the research system under study. $P_{top\ 0.01\%}$ can be calculated using the equation:

$$P_{top\ 0.01\%} = P_{top\ 10\%}\ (P_{top\ 1\%}/\ P_{top\ 10\%})^3 \qquad (1)$$

There are two ways to calculate $P_{top\ 0.01\%}$: (i) by calculating the parameter for each university and then summing the results, or (ii) by first summing $P_{top\ 10\%}$ and $P_{top\ 1\%}$ across all universities and then calculating $P_{top\ 0.01\%}$. The former method is more accurate, but country-level publication searches in bibliographic databases do not distinguish among institutions. Therefore, in this study, calculations were performed using the latter method unless otherwise indicated.

In addition to the *Leiden Ranking* data, I investigated five high-technology topics—semiconductors, solar cells, lithium batteries, composite materials, and graphene—and two biomedical topics—immunity and stem cells—using the same three indicators: $P_{top\ 10\%}$, $P_{top\ 1\%}/P_{top\ 10\%}$, and $P_{top\ 0.01\%}$. Much of the dataset was derived from the raw data of a previous publication (Rodríguez-Navarro & Brito, 2024b). In all cases, the searches were performed for ‘articles’ as described in that study using the Science Citation Index Expanded database in Clarivate's Web of Science Core Collection and the Advanced Search tool. The Citation Report tool was used to retrieve both publication counts and annual citation counts. Citation windows were delayed as much as possible for the reasons discussed in (Wang et al., 2017). For the 2004–2007 and 2014–2017 publication windows, the corresponding citation windows were 2009–2012 and 2019–2022, respectively. All reported data correspond to domestic papers in order to avoid the effect of USA–EU collaborations (Rodríguez-Navarro & Brito, 2024b).

It is worth noting that, in the *Leiden Ranking*, the proximity of the citation window to the publication period introduces a bias against novelty (Wang et al., 2017). The same problem applies to more recent publication years (2018, 2020, and 2022). In these cases, the citation windows were 2020–2021, 2022–2023, and 2024–2025, respectively.

The $e_p$ index used in a previous paper (Rodríguez-Navarro & Brito, 2018) is equivalent to the $P_{top\ 1\%}/P_{top\ 10\%}$ ratio. The only difference is statistical. The $e_p$ index is calculated by fitting a power law to at least five data points, whereas the $P_{top\ 1\%}/P_{top\ 10\%}$ ratio is calculated from only two points that belong to the power law distribution.

In the second part of this study, I used double-rank plots—global and country ranks—of the most cited papers. The information provided by these double-rank plots can be transformed into a numerical indicator using mathematical formulas (Rodríguez-Navarro, 2026; Rodríguez-Navarro & Brito, 2024b). However, in this study I use scatter plots because they are more informative visually. The relative positions of the countries' data points and their upward trends allow comparison of research activity at the breakthrough level.

It is important to keep in mind that the link between a publication's scientific importance and the number of citations is a correlation; therefore, the most cited paper is not necessarily groundbreaking—it could be a review paper classified as ‘article’ by the database or a publication describing a method. For this reason, it is necessary to consider several publications before making a judgment.

## 4. Results

### *4.1. EU, US, and Chinese universities*

The data depicted in Fig. 3 describe the characteristics of the top 10% most cited papers in the university systems of the EU, the USA, and China across the 14 evaluations of the *Leiden*

*Ranking* from 2011 to 2024. Over this period, the EU and US $P_{top\ 10\%}$ remained almost constant, with a slight advantage for the USA at the beginning of the period and for the EU at the end. Thus, high-quality 'normal' research was very similar in the EU and the USA throughout the entire period. In contrast to this stability in the EU and the USA, China's $P_{top\ 10\%}$ increased 7.5-fold. It started at a significantly lower level and ended at a significantly higher level than those of both the EU and the USA.

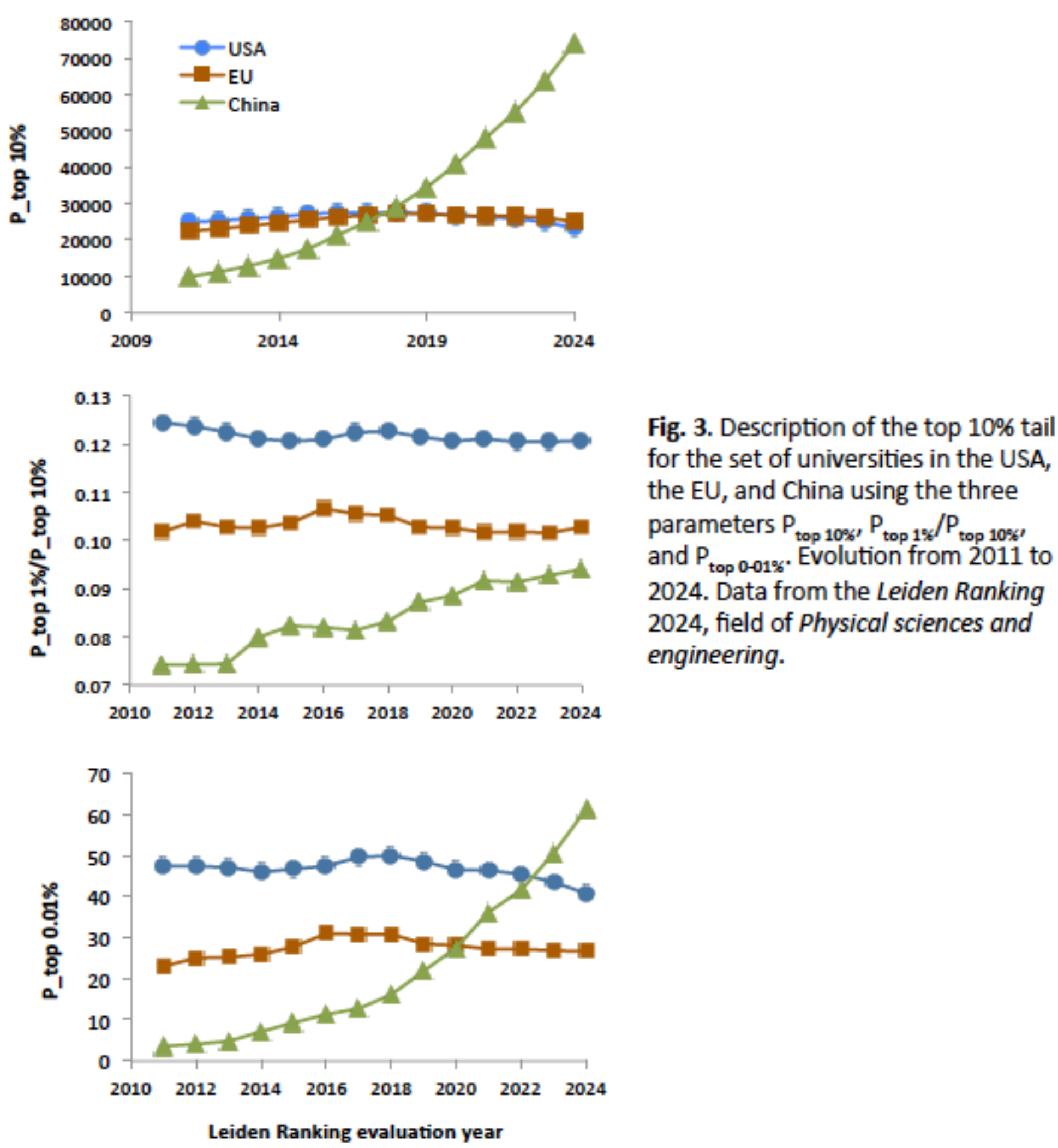

**Fig. 3.** Description of the top 10% tail for the set of universities in the USA, the EU, and China using the three parameters $P_{top\ 10\%}$, $P_{top\ 1\%}/P_{top\ 10\%}$, and $P_{top\ 0\text{-}01\%}$. Evolution from 2011 to 2024. Data from the *Leiden Ranking* 2024, field of *Physical sciences and engineering*.

Regarding the efficiency of research, as measured by the $P_{top\ 1\%}/P_{top\ 10\%}$ ratio, the conclusion is similar: very little changed in either the EU or the USA, with the USA maintaining a notable advantage over the EU (approximately 0.12 versus 0.10). In contrast, China's $P_{top\ 1\%}/P_{top\ 10\%}$ ratio improved from 0.07 to 0.09 and is currently close to that of the EU.

As a consequence of the observed changes in $P_{top\ 10\%}$ and $P_{top\ 1\%}/P_{top\ 10\%}$, $P_{top\ 0.01\%}$ varied very little in the USA and the EU, with US values remaining 50–60% higher than those of the EU. In contrast, China's $P_{top\ 0.01\%}$ increased 17-fold. Consequently, China surpassed the EU in 2021 and the USA in 2023. In 2024, China's $P_{top\ 0.01\%}$ was 2.3 times higher than that of the EU. In summary, Fig. 3 shows that high-quality 'normal' research, as measured by $P_{top\ 10\%}$, is very similar in the EU and the USA. The major difference appears in breakthrough research, as reflected by $P_{top\ 0.01\%}$.

When the $P_{top\ 0.01\%}$ values of individual universities are examined, the Leiden Ranking reveals an enormous difference between the leading universities in the USA and the EU. In 2024, the three US universities with the highest $P_{top\ 0.01\%}$ values—Massachusetts Institute of Technology, Harvard University, and Stanford University—had values ranging from 3.9 to 3.0. In contrast, the corresponding values for the three leading EU universities—University of Groningen, Delft University of Technology, and Carl von Ossietzky University of Oldenburg—ranged from 0.8 to 0.7. To illustrate these differences more clearly, Fig. 4 shows the cumulative $P_{top\ 0.01\%}$ values of US and EU universities after ordering them from the highest to the lowest $P_{top\ 0.01\%}$ value. These cumulative values differ from the country values shown in Fig. 3 (34.2 versus 26.7 for the EU and 48.1 versus 40.8 for the USA). This difference arises from the different methods used to calculate $P_{top\ 0.01\%}$, as explained in Section 3. Nevertheless, both Figs. 3 and 4 demonstrate the dominance of the USA in breakthrough research.

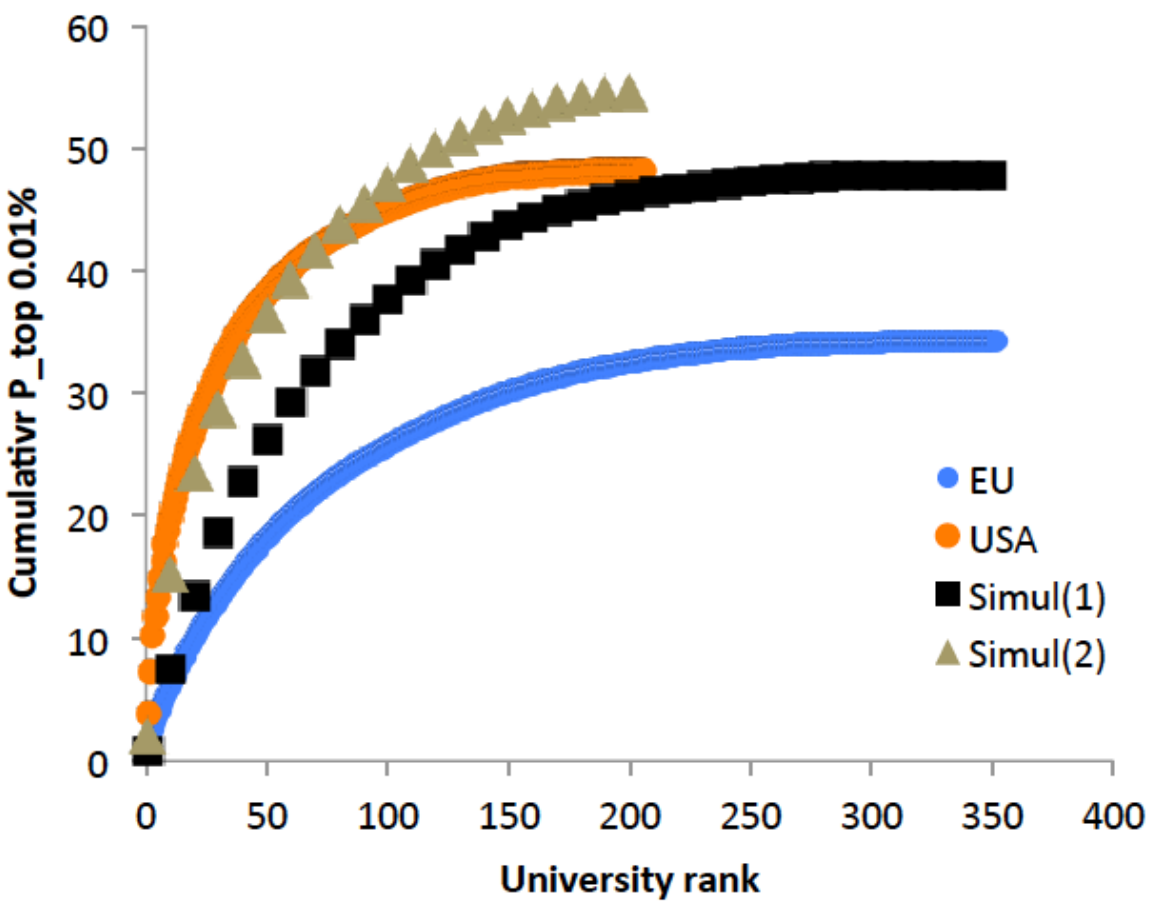

**Fig. 4**. Cumulative plot of $P_{top\ 0.01\%}$ in EU and US universities ordered from the highest to the lowest values of the parameter. Data from *Leiden Ranking* 2024. In Simul(1), the proportion of universities matching the best Dutch universities was increased; in Simul(2), UK universities were repeated four times.

The conjecture that, to be competitive, the EU needs universities comparable to the highest-ranked US universities in terms of $P_{top\ 0.01\%}$ is incorrect. It would also be unrealistic because the federal structure of the USA differs substantially from the political structure of the EU. In fact, the EU could achieve cumulative $P_{top\ 0.01\%}$ values equal to or higher than those of the USA without having universities at the level of the top US institutions. To illustrate this point, Fig. 4 includes two simulations. In the first simulation: Simul(1), I used the observed data for EU universities but increased the proportion of universities with $P_{top\ 0.01\%}$ values between 0.8 and 0.3, while keeping the total number of universities unchanged. In the second simulation: Simul(2), I replicated the data for UK universities four times to obtain approximately the same number of universities as in the USA (200 versus 206). This latter simulation represents the hypothetical case in which the universities of Germany, France, Italy, and Spain performed at the same level as UK universities. The $P_{top\ 0.01\%}$ values for the three leading UK universities—Cambridge, Oxford, and Imperial College London—are 2.1, 1.3, and 0.9, respectively. In the first simulation, the EU would reach the same cumulative $P_{top\ 0.01\%}$ value as the USA. In the second simulation, with approximately the same number of universities as the USA, the simulated EU value would exceed that of the USA (54.6 versus 48.1). These simulations demonstrate that, by improving university research, the EU could match or surpass the USA without requiring universities comparable to the very top US institutions.

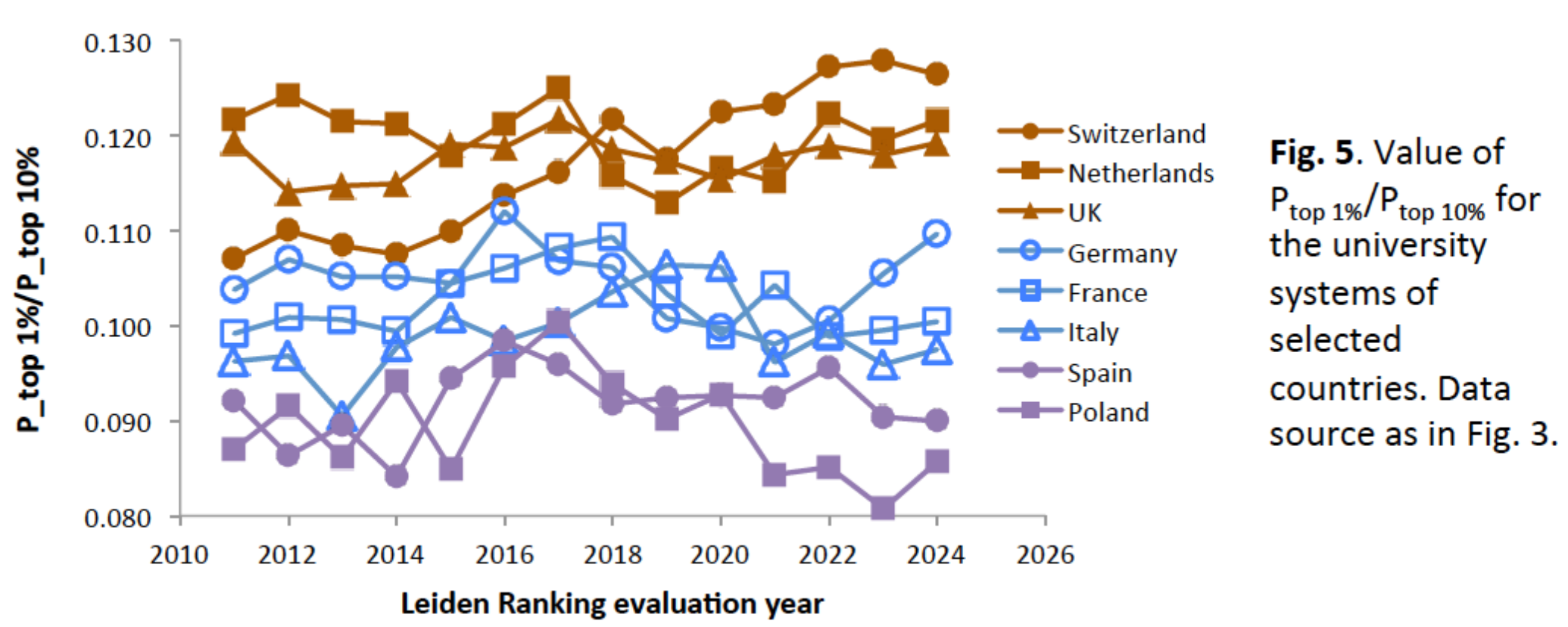

**Fig. 5**. Value of $P_{top\ 1\%}/P_{top\ 10\%}$ for the university systems of selected countries. Data source as in Fig. 3.

The simulations shown in Fig. 4 demonstrate that improving breakthrough research in the EU is entirely feasible if the member states improve their research systems. Certainly, increasing $P_{top\ 10\%}$ through greater investment would also increase $P_{top\ 0.01\%}$ (Eq. 1). However, this approach would require unacceptably large investments. The only realistic research policy for the EU is therefore to increase the $P_{top\ 1\%}/P_{top\ 10\%}$ ratio. Unfortunately, this is not occurring. Fig. 5 shows the evolution of this ratio over the 14-year study period for the five largest European countries—Germany, France, Italy, Spain, and Poland—compared with the Netherlands, Switzerland, and the UK. Based on the $P_{top\ 1\%}/P_{top\ 10\%}$ ratio, three groups of countries can be distinguished: the leaders (the Netherlands, Switzerland, and the UK), followed by Germany, France, and Italy, and finally Spain and Poland. The important point is that, over the 14-year period recorded in Fig. 5, there has been no tendency for the countries in the latter two groups to reduce the gap with the Netherlands, Switzerland, and the UK.

One might argue that countries are constrained to a certain research level because of structural factors and that changing this level is difficult or even impossible. Figure 6 shows that this is not the case by comparing Australia with France and Singapore with Norway. In France, $P_{top\ 0.01\%}$ changed very little throughout the study period, whereas in Australia it increased fourfold. It began at a value 70% lower than that of France, reached parity in 2018, and by 2023–2024 Australia's value was approximately three times higher than France's.

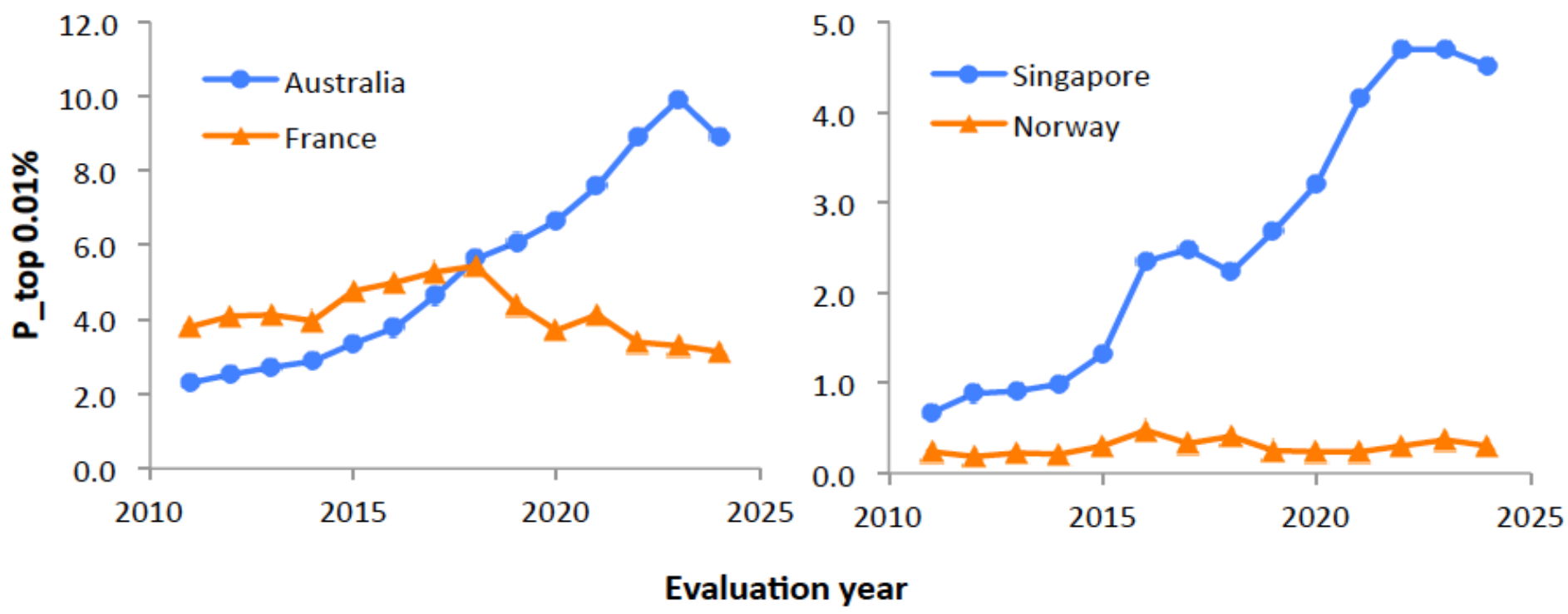

**Fig. 6**. Evolution of $P_{top\ 0.01\%}$ for the university systems of Australia, France, Singapore, and Norway. Data source as in Fig. 3.

Particularly interesting is the comparison between Norway and Singapore, two countries with similarly wealthy economies and comparable population sizes. During the study period, Singapore increased its $P_{top\ 0.01\%}$ approximately sevenfold, whereas Norway's value remained almost unchanged. It is worth noting that the GDP per capita of both Singapore and Norway is very high—approximately twice that of France—which implies that a low GDP per capita, an impediment to efficient research (Rodríguez-Navarro & Brito, 2022), cannot explain Norway's poorer research performance relative to Singapore.

*4.2. EU, USA, and China in high-tech topics*

Using the *Leiden Ranking* data, $P_{top\ 0.01\%}$, calculated from $P_{top\ 10\%}$ and $P_{top\ 1\%}$ as an average across many research fields, is an indicator of the capacity of university research to produce breakthrough results, but it is not a measure of performance in specific high-technology topics. To assess such performance, Table 1 reports *P*, $P_{top\ 10\%}$, $P_{top\ 1\%}/P_{top\ 10\%}$, and $P_{top\ 0.01\%}$ for four technological topics—semiconductors, solar cells, lithium batteries, and composite materials—for the 2014–2017 and 2021–2022 publication periods.

**Table 1.** Evaluation parameters for publications from the USA, China, and the EU in the four indicated high-tech topics and the indicated publication periods.[a]

| | 2014–2017 (4 years) | | | | | 2021–2022 (2 years) | | | |
|---|---|---|---|---|---|---|---|---|---|
| | **World** | **USA** | **China** | **EU** | | **World** | **USA** | **China** | **EU** |
| **Semiconductors** | | | | | | | | | |
| P | 58526 | 6019 | 12183 | 7452 | | 38812 | 2437 | 11741 | 3522 |
| $P_{top\ 10\%}$ | | 511 | 1353 | 511 | | | 232 | 1463 | 249 |
| $P_{top\ 1\%}$ | | 100 | 102 | 35 | | | 24 | 135 | 18 |
| $P_{top\ 1\%}/P_{top\ 10\%}$ | | 0.20 | 0.08 | 0.07 | | | 0.10 | 0.09 | 0.07 |
| $P_{top\ 0.01\%}$ | | 3.83 | 0.58 | 0.16 | | | 0.26 | 1.15 | 0.09 |
| **Solar cells** | | | | | | | | | |
| P | 61699 | 5505 | 12883 | 8220 | | 43910 | 1773 | 11053 | 4649 |
| $P_{top\ 10\%}$ | | 773 | 1116 | 653 | | | 175 | 1362 | 328 |
| $P_{top\ 1\%}$ | | 113 | 73 | 32 | | | 24 | 125 | 32 |
| $P_{top\ 1\%}/P_{top\ 10\%}$ | | 0.15 | 0.07 | 0.05 | | | 0.14 | 0.09 | 0.10 |
| $P_{top\ 0.01\%}$ | | 2.41 | 0.31 | 0.08 | | | 0.45 | 1.05 | 0.30 |
| **Lithium batteries** | | | | | | | | | |
| P | 32354 | 3376 | 12577 | 2972 | | 32012 | 1984 | 15597 | 2503 |
| $P_{top\ 10\%}$ | | 584 | 983 | 244 | | | 254 | 1471 | 154 |
| $P_{top\ 1\%}$ | | 96 | 83 | 21 | | | 30 | 114 | 10 |
| $P_{top\ 1\%}/P_{top\ 10\%}$ | | 0.16 | 0.08 | 0.09 | | | 0.12 | 0.08 | 0.06 |
| $P_{top\ 0.01\%}$ | | 2.59 | 0.59 | 0.16 | | | 0.42 | 0.68 | 0.04 |
| **Composite materials** | | | | | | | | | |
| P | 50260 | 4011 | 11731 | 8831 | | 50446 | 2510 | 15119 | 6725 |
| $P_{top\ 10\%}$ | | 437 | 1285 | 653 | | | 301 | 1829 | 466 |
| $P_{top\ 1\%}$ | | 75 | 122 | 46 | | | 37 | 201 | 23 |
| $P_{top\ 1\%}/P_{top\ 10\%}$ | | 0.17 | 0.09 | 0.07 | | | 0.12 | 0.11 | 0.05 |
| $P_{top\ 0.01\%}$ | | 2.21 | 1.10 | 0.23 | | | 0.56 | 2.43 | 0.06 |
| **Combination of the four technologies** | | | | | | | | | |
| P | 202839 | 18911 | 49374 | 27475 | | 165180 | 8704 | 53510 | 17399 |
| $P_{top\ 10\%}$ | | 2305 | 4737 | 2061 | | | 962 | 6125 | 1197 |
| $P_{top\ 1\%}$ | | 384 | 380 | 134 | | | 115 | 575 | 83 |
| $P_{top\ 1\%}/P_{top\ 10\%}$ | | 0.17 | 0.08 | 0.07 | | | 0.12 | 0.09 | 0.07 |
| $P_{top\ 0.01\%}$ | | 10.66 | 2.45 | 0.57 | | | 1.64 | 5.07 | 0.40 |

[a] P, total number of publications, $P_{top\ x\%}$, number of publications in top x% by citation. Citation windows described in text

For the 2014–2017 period, the combined data for the four technologies (Table 1, "Combination of the four technologies") are broadly consistent with those obtained from the *Leiden Ranking* for a comparable period, although some differences exist. The most important difference is that the US $P_{top\ 1\%}/P_{top\ 10\%}$ ratio is considerably higher than that obtained from the *Leiden Ranking* (0.17 versus 0.12), whereas the opposite is true for the EU (0.07 versus 0.10). For China, the ratios are identical. With respect to $P_{top\ 0.01\%}$, the USA clearly dominated the four technologies, with values approximately fourfold higher than China's and 20-fold higher than the EU's.

The 2021–2022 period showed a dramatic change compared with the 2014–2017 period (after correcting for the difference between four-year and two-year periods), particularly in the number of publications from China. Whereas *P* and $P_{top\ 10\%}$ changed very little in the USA and the EU—they decreased slightly in the USA and increased slightly in the EU—they increased more than twofold in China. Although China's $P_{top\ 1\%}/P_{top\ 10\%}$ ratio improved only marginally, the dramatic increase in $P_{top\ 10\%}$ resulted in China overtaking the USA in terms of $P_{top\ 0.01\%}$ (5.1 versus 1.6).

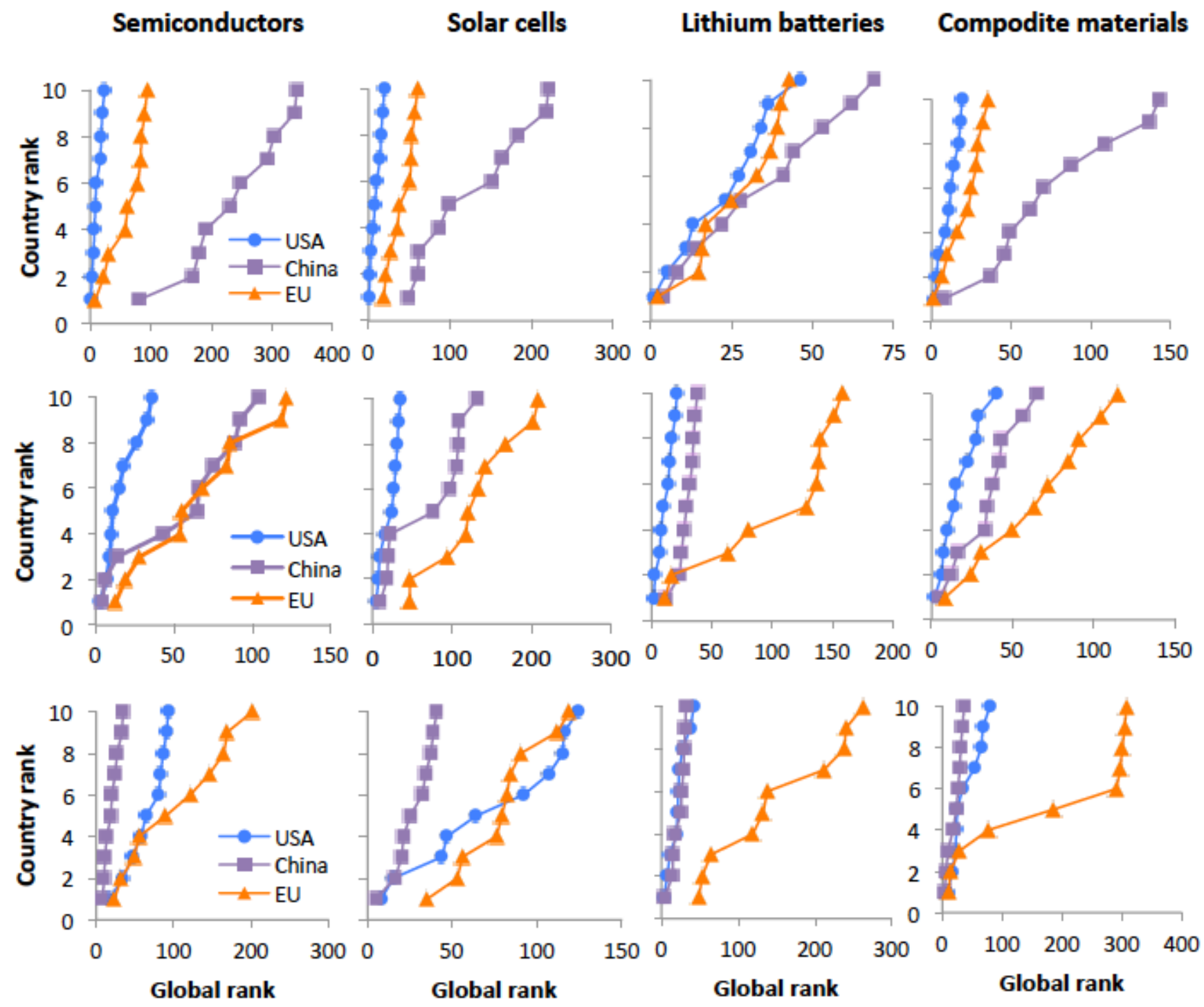

**Fig. 7**. Double-rank plots of the 10 most cited domestic papers from the USA, China, and the EU in the four indicated high-tech topics. Three evaluation periods: 2004–2007 (first panel row), 2014–2017 (second panel row), and 2021–2022 (third panel row). Citation windows: 2009–2012, 2019–2022, and 2024–2025, respectively.

Table 1 reflects research quality as inferred from the top 10% tail of the most cited papers. However, this approach is almost insensitive to the possible existence of elite groups of researchers who, although small in number, publish papers that receive substantially more citations than would be expected from the overall citation distribution (Rodríguez-Navarro, 2024). The study of the most cited papers without prior assumptions about the citation distribution can be performed using several indicators based on the global ranks of the most cited papers in each country (Rodríguez-Navarro & Brito, 2024b). However, a more visually intuitive representation is obtained by plotting country rank against the global ranks of the most cited papers. Figure 7 presents the double-rank plots for the 10 most cited papers from the USA, China, and the EU in the four technologies included in Table 1 for the periods 2004–2007, 2014–2017, and 2021–2022—it should be noted that the success of a country as measured by the global ranks of the most cited papers is size-dependent.

As a general rule, during the 2004–2007 period, US dominance over the EU was clear, except in lithium batteries, whereas China was still a weak competitor. During the 2014–2017 period, the USA maintained its leading position, but China overtook the EU and clearly established itself in second place. The interpretation of the 2021–2022 results is somewhat more uncertain because the citation window is not optimal for identifying groundbreaking papers (Section 3). Nevertheless, the observed patterns are very clear, and it is unlikely that the inappropriate citation window has introduced substantial bias. During this period, China emerged as the unquestionable leader in all four technologies except lithium batteries, where the USA and China alternated among the top 10 global ranks. By 2021–2022, the EU's contribution to breakthrough research in these technologies had become negligible compared with China's.

### *4.3. Graphene, a benchmarking topic*

The study of graphene research provides an excellent benchmark for investigating how national research systems adapt to emerging technologies. First, intensive research on graphene began only recently, following the publication of the pioneering and groundbreaking paper by Novoselov *et al.* (2004), and expanded rapidly after the awarding of the 2010 Nobel Prize in Physics to Andre Geim and Konstantin Novoselov. Second, graphene is a material with numerous applications in next-generation technologies (Singh et al., 2025). The EU responded to this technological challenge by launching the Graphene Flagship (https://graphene-flagship.eu/about/first-10-years, accessed on June, 2026).

Table 2 presents the same indicators for graphene as those reported in Table 1 for the other technologies, and Fig. 8 shows the corresponding double-rank plots for the USA, China, and the EU. According to $P_{top\ 0.01\%}$, the USA was the clear leader during the 2014–2017 period. However, the double-rank plots in Fig. 8 suggest that this leadership was less definitive. By 2018, China's dominance had become evident according to both $P_{top\ 0.01\%}$ (Table 2) and the double-rank plots (Fig. 8), and by 2020 and 2022 China's leadership was unequivocal.

**Table 2.** Evaluation parameters for publications from the USA, China, and the EU in graphene during the indicated periods.[a]

| 2014-2017 | | | | |
|---|---|---|---|---|
| | World | USA | China | EU |
| P | 82839 | 4677 | 32745 | 5854 |
| $P_{top\ 10\%}$ | | 603 | 3350 | 321 |
| $P_{top\ 1\%}$ | | 121 | 252 | 28 |
| $P_{top\ 1\%}/P_{top\ 10\%}$ | | 0.20 | 0.08 | 0.09 |
| $P_{top\ 0.01\%}$ | | 4.87 | 1.43 | 0.21 |
| **2018** | | | | |
| P | 30037 | 1137 | 12774 | 1713 |
| $P_{top\ 10\%}$ | | 120 | 1474 | 73 |
| $P_{top\ 1\%}$ | | 20 | 122 | 4 |
| $P_{top\ 1\%}/P_{top\ 10\%}$ | | 0.17 | 0.08 | 0.05 |
| $P_{top\ 0.01\%}$ | | 0.56 | 0.84 | 0.01 |
| **2020** | | | | |
| P | 34494 | 1050 | 14175 | 1861 |
| $P_{top\ 10\%}$ | | 111 | 1629 | 74 |
| $P_{top\ 1\%}$ | | 13 | 143 | 4 |
| $P_{top\ 1\%}/P_{top\ 10\%}$ | | 0.12 | 0.09 | 0.05 |
| $P_{top\ 0.01\%}$ | | 0.18 | 1.10 | 0.01 |
| **2022** | | | | |
| P | 38467 | 947 | 16728 | 1890 |
| $P_{top\ 10\%}$ | | 72 | 1978 | 98 |
| $P_{top\ 1\%}$ | | 12 | 193 | 7 |
| $P_{top\ 1\%}/P_{top\ 10\%}$ | | 0.17 | 0.10 | 0.07 |
| $P_{top\ 0.01\%}$ | | 0.33 | 1.84 | 0.04 |

[a] P, total number of publications, $P_{top\ x\%}$, number of publications in top x% by citation. Citation windows described in text

China's leadership over the USA in graphene research has the same underlying components as in the other technologies (Table 1). According to the $P_{top\ 1\%}/P_{top\ 10\%}$ ratio, the US research system, with a value of approximately 0.17, is considerably more efficient than China's, whose ratio is approximately 0.09. The major difference between China and the USA, however, lies in $P_{top\ 10\%}$, which in 2022 was 27-fold higher in China than in the USA.

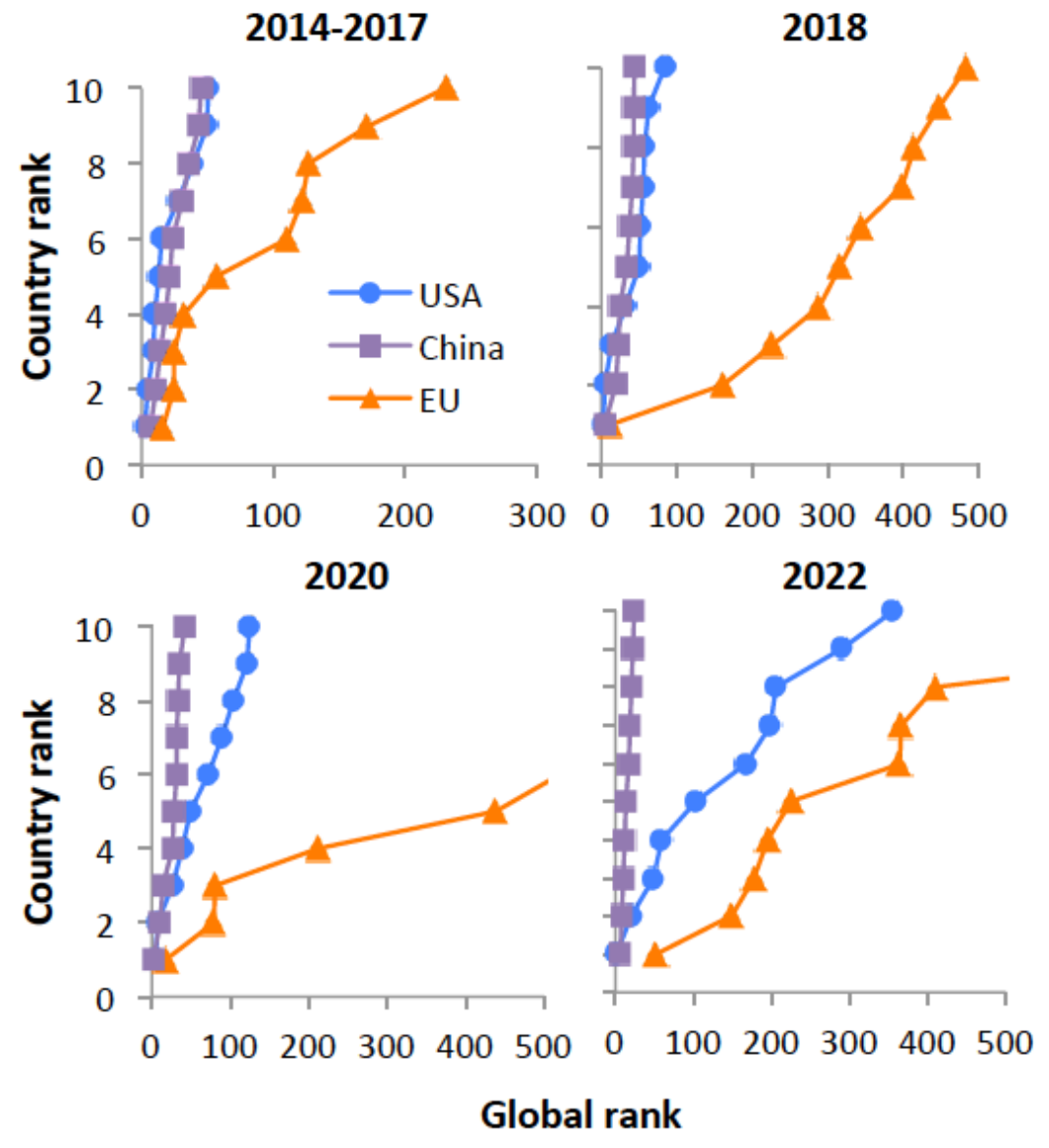

**Fig. 8**. Double rank plots of the 10 most cited domestic papers from the USA, China, and the EU in graphene. Citation windows: 2019–2022, 2020–2022, 2022–2023, and 2024–2025, respectively.

In the EU, breakthrough research on graphene remains of limited significance compared with that of China and the USA. As in the other technologies, this reflects both a much lower $P_{top\ 1\%}/P_{top\ 10\%}$ ratio than that of the USA and a much lower $P_{top\ 10\%}$ than that of China.

*4.4. EU, USA, and China in in biomedicine*

Although the EU's competitiveness gap in the pharmaceutical sector (Draghi, 2024b) deserves to be studied, assessing research in this area presents specific methodological difficulties because, in biomedical fields, clinical guidelines and the results of clinical trials—which are typically very highly cited—interfere with the analysis of breakthrough research.

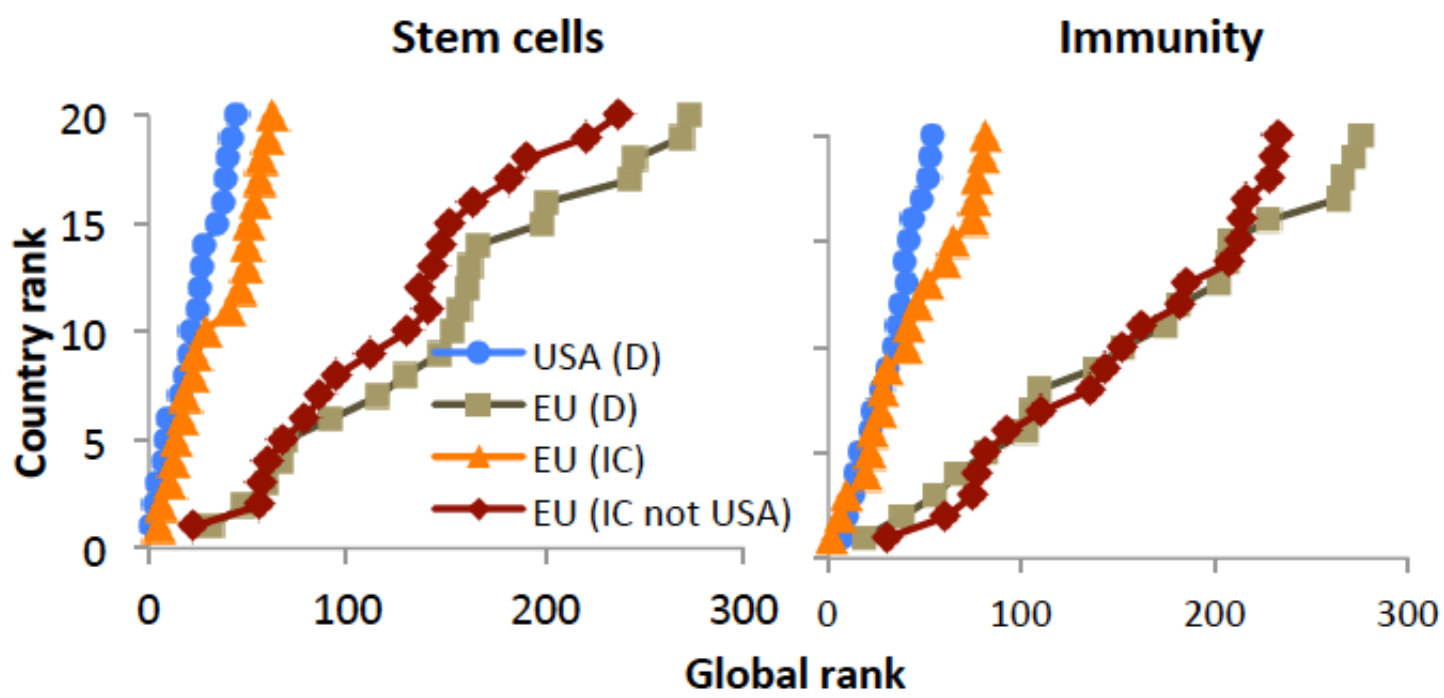

**Fig. 9.** Effect of US collaborations with the EU in the research topics of stem cells and immunity. Double-rank plots of EU papers: domestic papers (D), international collaborations (IC), and international collaborations excluding the USA. Domestic US papers are shown as the reference. Publication window, 2014–2017; citation window, 2019–2022.

In the *Leiden Ranking*, an analysis of *Biomedical and Health Sciences* similar to that shown in Fig. 3 presents an additional difficulty. In biomedical fields, the overwhelming dominance of the USA (Rodríguez-Navarro, 2025b) introduces a strong bias when either full or fractional counting is used. The double-rank plots shown in Fig. 9 illustrate how EU–US

collaborations—but not collaborations with other countries—improve the apparent performance of EU research. International collaborations substantially improve the global ranks of the EU's most cited papers relative to domestic papers, bringing them almost to the level of US domestic papers. However, this improvement largely disappears when collaborations with the USA are excluded.

**Table 3**. Evaluation parameters for publications from the USA, China, and the EU in immunity and stem cells during the indicated publication periods.[a]

| | 2014–2017 (4 years) | | | | | 2021–2022 (2 years) | | | |
|---|---|---|---|---|---|---|---|---|---|
| | World | USA | China | EU | | World | USA | China | EU |
| **Immunity** | | | | | | | | | |
| P | 42798 | 7963 | 6752 | 6694 | | 38223 | 5015 | 10970 | 4810 |
| $P_{top\ 10\%}$ | | 1156 | 370 | 510 | | | 631 | 1050 | 332 |
| $P_{top\ 1\%}$ | | 141 | 16 | 28 | | | 72 | 91 | 15 |
| $P_{top\ 1\%}/P_{top\ 10\%}$ | | 0.12 | 0.04 | 0.05 | | | 0.11 | 0.09 | 0.05 |
| $P_{top\ 0.01\%}$ | | 2.10 | 0.03 | 0.08 | | | 0.94 | 0.68 | 0.03 |
| **Stem cells** | | | | | | | | | |
| P | 86876 | 16464 | 14713 | 13931 | | 51230 | 6785 | 13703 | 7095 |
| $P_{top\ 10\%}$ | | 2435 | 1179 | 1240 | | | 741 | 1605 | 383 |
| $P_{top\ 1\%}$ | | 272 | 46 | 63 | | | 86 | 129 | 24 |
| $P_{top\ 1\%}/P_{top\ 10\%}$ | | 0.11 | 0.04 | 0.05 | | | 0.12 | 0.08 | 0.06 |
| $P_{top\ 0.01\%}$ | | 3.39 | 0.07 | 0.16 | | | 1.16 | 0.83 | 0.09 |

[a] P, total number of publications, $P_{top\ x\%}$, number of publications in top x% by citation. Citation windows described in text

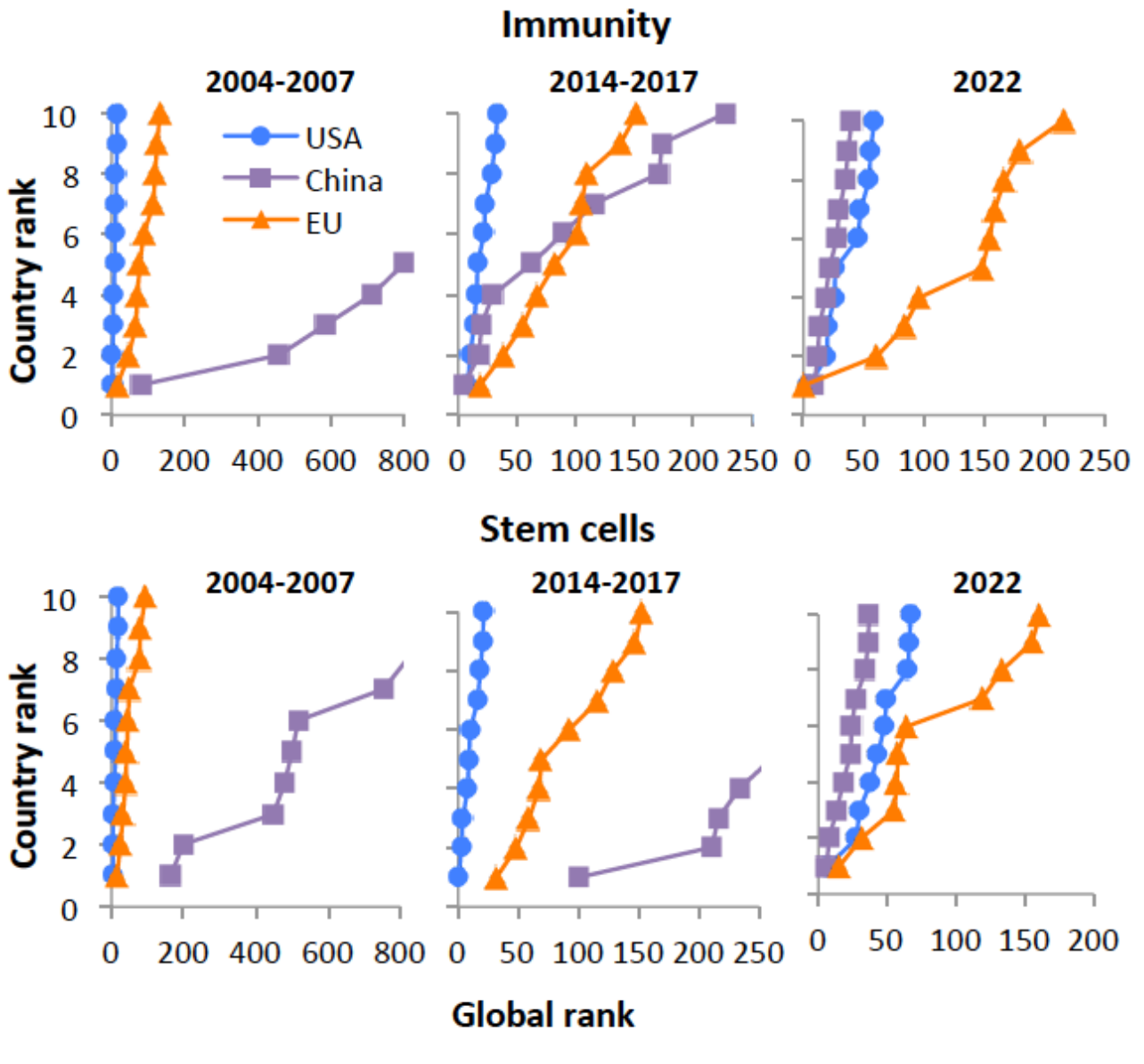

**Fig. 10.** Double-rank plots of the 10 most cited domestic papers from the USA, China, and the EU in immunity and stem cells. Citation windows: 2009–2012, 2019–2022, and 2024–2025, respectively.

To obtain an overall view of research performance in biomedical fields, I investigated two research topics—immunity and stem cells—in which the influence of clinical studies is less pronounced. Table 3 and Fig. 10 present the results, which lead to conclusions similar to those obtained for the technological topics, although China's progress in these two biomedical fields is even more remarkable. The clear US leadership during the 2014–2017 period was still evident in the 2021–2022 period according to Table 3, but by 2022 China had overtaken the USA according to the double-rank plots shown in Fig. 10. Once again, as in the technological topics, the EU's contribution to breakthrough research in these two biomedical fields is almost negligible.

### *4.5. EU countries fail in the frontier of knowledge*

Pushing the boundaries of knowledge depends on both efficiency and size (Section 1.4). Consequently, the size of individual EU countries is generally insufficient for them to compete with the USA or China. However, the EU's difficulty in competing with the USA, despite their comparable overall size, raises the question of which EU countries contribute most to the EU's performance at the frontier of knowledge. Although this question has already been addressed using quantitative indicators in previous studies (Rodríguez-Navarro & Brito, 2018, 2024b), here I examine it from a different perspective.

A notable characteristic of disruptive innovation is that, in many cases, success belongs only to the first to achieve the scientific advance underlying the innovation (Section 1.4). Although, as noted above, it is unlikely that any single EU country will produce papers occupying the very top positions in the global citation rankings, examining the rankings of the most cited papers from individual countries can still provide valuable information about their research performance. To this end, Table 4 reports the global ranks of the three most cited papers from the largest EU countries in the topics already examined. For comparison, the table also includes the EU as a whole, the USA, and the UK. Country size is represented by GDP.

As expected, the variability of the rankings shown in Table 4 is high. However, if the UK rankings in semiconductors, solar cells, and composite materials are taken as benchmarks for a country that is competitive at the frontier of knowledge, only France, in composite materials, can be considered competitive in the corresponding research field. The total number of papers in the studied topics and publication periods ranges from approximately 50,000 to 80,000, implying that even a paper ranked 1000 is within the top 2% globally. Therefore, all the papers listed in Table 4 are of high quality, but they are not the breakthrough papers that push the boundaries of knowledge. Compared with the USA, the EU countries remain trapped in 'normal' research: the UK because of its relatively small size, and the other countries because they combine limited size with relatively low research efficiency.

**Table 4**. Global ranks of the three most cited papers in the EU and indicated countries in the seven stated topics in the period 2014-2017.

| Country | GDP | Semiconductors | Solar cells | Lithium batteries | Composite materials | Graphene | Immunity | Stem cells |
|---|---|---|---|---|---|---|---|---|
| EU | 21.3 | 13<br>19<br>28 | 46<br>48<br>95 | 11<br>16<br>64 | 8<br>24<br>31 | 16<br>24<br>25 | 19<br>38<br>55 | 32<br>48<br>59 |
| Germany | 5.5 | 55<br>131<br>132 | 95<br>251<br>321 | 81<br>128<br>139 | 91<br>167<br>184 | 244<br>261<br>567 | 38<br>55<br>208 | 32<br>166<br>243 |
| France | 3.6 | 86<br>359<br>391 | 453<br>907<br>1081 | 256<br>281<br>315 | 8<br>50<br>63 | 24<br>171<br>833 | 174<br>366<br>639 | 302<br>759<br>900 |
| Italy | 2.7 | 19<br>119<br>199 | 48<br>118<br>203 | 782<br>951<br>1292 | 85<br>135<br>263 | 1007<br>1890<br>2073 | 102<br>328<br>538 | 92<br>162<br>381 |
| Spain | 2.1 | 28<br>200<br>420 | 167<br>404<br>574 | 295<br>1092<br>2454 | 31<br>306<br>329 | 404<br>442<br>596 | 510<br>980<br>1068 | 59<br>116<br>273 |
| USA | 32.4 | 3<br>7<br>9 | 4<br>8<br>9 | 2<br>3<br>7 | 2<br>6<br>7 | 3<br>5<br>9 | 8<br>10<br>13 | 1<br>3<br>4 |
| UK | 4.3 | 30<br>61<br>63 | 16<br>25<br>37 | 78<br>287<br>341 | 23<br>55<br>74 | 44<br>468<br>577 | 68<br>78<br>96 | 125<br>135<br>246 |

### *4.6. The dominance of China*

Overall, the results presented above indicate that China caught up with the USA in breakthrough research during the 2020–2022 period—earlier in the case of graphene—which raises the question of whether China has continued to extend its lead. The citation window required to calculate $P_{top\ 0.01\%}$ does not allow the current position of China to be assessed in recent years. However, the relationship among $P$, $P_{top\ 10\%}$, and $P_{top\ 1\%}$ has remained fairly stable in both the USA and China during the years examined. This suggests that a reasonable estimate of $P_{top\ 0.01\%}$ can be inferred from the number of published papers. Using the data reported in Fig. 11, the widening gap between the two countries over the 2020–2025 period suggests that China's advantage observed in 2022 has probably continued to increase during the subsequent three years. More importantly, because the growth in China's publication output had not slowed by 2025, it is reasonable to predict that, within a few more years, most breakthrough research in high-technology sectors will be produced in China unless the USA and the EU significantly improve their research performance.

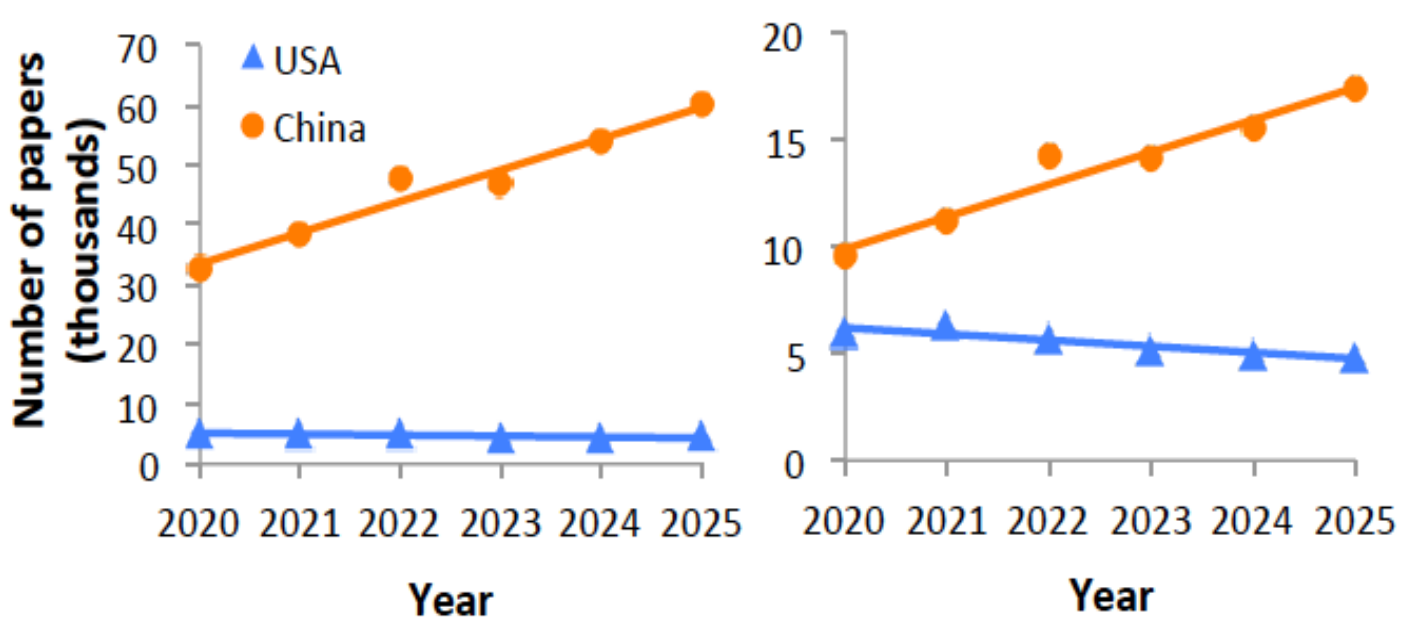

**Fig. 11.** Number of domestic publications per year from the USA and China in five technological topics (left): semiconductors, solar cells, lithium batteries, composite materials, and graphene; and two biomedical topics (right): immunity and stem cells.

## 5. Discussion

Compared with the USA and China, economic analyses describe the EU as having a weak capacity to compete in terms of disruptive innovation (Draghi, 2024a), and the EU industry and economy appear to be trapped in a middle-technology trap (Fuest et al., 2024). The causes most frequently cited to explain this situation are the existence of an "investment gap" in innovation (Draghi, 2024a; Letta, 2024) and difficulties in transferring research to the market (Section 1.1). Gros et al. (2025) argue that the "investment gap" is actually an "R&D gap."

In line with the idea of an "R&D gap," this study provides an explanation based on previous publications describing a gap in breakthrough research, whereas no such gap—or only a small one—exists in high-quality 'normal' research. Indeed, in almost all technological topics, the EU publishes a similar or even larger number of papers than the USA, and the differences in $P_{top\ 10\%}$ are not substantial. However, despite these relatively small differences, $P_{top\ 0.01\%}$ and the rankings of the most highly cited papers demonstrate that the EU does not compete with the USA and China in breakthrough research (Tables 1–3; Figs. 7, 8, and 10; see also Rodríguez-Navarro, 2025b; Rodríguez-Navarro & Brito, 2018).

This gap in breakthrough research is probably the main cause of the EU investment gap in innovation. Because of this gap, the EU is less attractive to investors in high-tech sectors (Section 1.4). Given China's current growth, it is difficult to envision a solution to the EU's deficit in high-tech innovation. However, if such a solution exists, it would require a revision of the research policies pursued over the past 20–30 years.

### *5.1. The EU breakthrough research gap*

Assuming that the global frequency of breakthrough papers is approximately one per 10,000 ‘normal’ papers (Bornmann et al., 2018; Poege et al., 2019), breakthrough research cannot be evaluated simply by counting papers. To address this problem, this study uses two approaches: (i) calculation of the $P_{top\ 1\%}/P_{top\ 10\%}$ ratio—a size-independent indicator of research efficiency—and $P_{top\ 0.01\%}$—a size-dependent indicator of the number of breakthrough papers—taking advantage of the power-law distribution of percentile papers (Section 3); and (ii) analysis of the double-rank plots of the most highly cited papers.

The first approach was applied to *Leiden Ranking* data in the field of *Physical Sciences and Engineering* and to several high-tech topics in which innovation depends on advances at the knowledge frontier. The results clearly show that EU research is much less competitive than US research, and this difference is even more pronounced in high-tech topics (Fig. 3 and Table 1). For example, the $P_{top\ 1\%}/P_{top\ 10\%}$ ratios calculated from the Leiden Ranking are approximately 0.10 for the EU and 0.12 for the USA, whereas in high-tech topics they are approximately 0.07 for the EU and 0.12 for the USA. Graphene research deserves special attention because of the unique characteristics of this field (Section 4.3). In this case, the difference between the USA and the EU is even larger, with $P_{top\ 1\%}/P_{top\ 10\%}$ ratios of approximately 0.17 and 0.07, respectively.

Although the EU publishes more papers than the USA, the differences in $P_{top\ 10\%}$ are much smaller, and in some cases $P_{top\ 10\%}$ is even higher in the USA (Fig. 1). Combined with the higher $P_{top\ 1\%}/P_{top\ 10\%}$ ratio in the USA, this results in $P_{top\ 0.01\%}$ values that are generally much higher in the USA than in the EU. In lithium batteries, composite materials, and graphene, the difference is approximately tenfold.

These results are consistent with those of a previous study based on the $e_p$ index, which is equivalent to the $P_{top\ 1\%}/P_{top\ 10\%}$ ratio (Section 3) and was calculated for fast- and slow-evolving technologies. In slow-evolving technologies, the $e_p$ index values for the EU and the USA are 0.09 and 0.12, respectively, whereas in fast-evolving technologies they are 0.06 and 0.13, respectively (Rodríguez-Navarro & Brito, 2018). This dependence on the type of technology explains why the differences observed using the *Leiden Ranking* are smaller than those observed in high-tech topics, since many of the microfields used in the Leiden Ranking are slow evolving.

The second approach, based on the double-rank plots of the 10 most highly cited papers, leads to conclusions regarding the US advantage over the EU that are similar to those obtained from $P_{top\ 0.01\%}$, or even less favorable for the EU (Fig. 7 and 8). Although the two methods do not produce identical results, they lead to the same conclusion: the EU's contribution to global breakthrough research is very limited.

The *Leiden Ranking* data (Fig. 3) show that China's progress in recent years has been spectacular, as described in previous studies (Tang et al., 2021; Wagner et al., 2022). The most notable feature is the enormous increase in $P_{top\ 0.01\%}$ (Fig. 3; Tables 1 and 2) and China's emergence as a global leader in the 2021–2022 period (Fig. 7 and 8), driven primarily by a massive increase in the number of publications. Remarkably, during the 2021–2022 period, domestic Chinese publications on lithium batteries accounted for nearly half of all papers published worldwide. In contrast, China's $P_{top\ 1\%}/P_{top\ 10\%}$ ratio remains modest, below the world average of 0.10. However, this result should be interpreted with caution. If there are "excessive citations from Chinese researchers to domestic papers" (Cai et al., 2019, p. 49), such citation practices may increase $P_{top\ 10\%}$ more than $P_{top\ 1\%}$, because the relative effect decreases as citation counts become very high, thereby reducing the $P_{top\ 1\%}/P_{top\ 10\%}$ ratio.

The above discussion focuses on technological fields, but the performance of the EU and the emergence of China in biomedical research have followed very similar patterns. Although these fields require specific analytical approaches (Section 4.4), the present data for two biomedical topics do not differ fundamentally from those for technological fields, with the only notable exception being that China's rise occurred later.

### *5.2. The EC overlooks the EU breakthrough research gap*

Although the EU innovation gap is widely recognized (Section 1.1), its relationship to the breakthrough research gap described in the previous section is frequently overlooked.

For example, the EC identified a weakness in EU research based on $P_{top\ 1\%}$ (European Commission, 2024a, Box 2), but this finding was largely overlooked and did not trigger any noticeable alarm. This lack of concern can be understood because the EU deficit in $P_{top\ 1\%}$, relative to China, is approximately 50%, and relative to the USA it is approximately 20% (European Commission, 2024b). However, in technological fields, these relatively small differences represent only the tip of the iceberg. First, in the EC reports, $P_{top\ 1\%}$ is calculated as a single universal indicator, and in this type of analysis, research in high-tech topics is diluted by the much larger volume of publications in fields with limited economic relevance. Second, the citation windows used in the EC reports—for example, 2020–2022 for publications published in 2020—introduce a strong bias against breakthrough papers (Wang et al., 2017).

Furthermore, $P_{top\ 1\%}$ is not an appropriate indicator for assessing breakthrough research, because breakthrough papers are much rarer. Consequently, a modest research deficit detected using $P_{top\ 1\%}$ translates into a much larger deficit in countries with low efficiency in generating breakthrough research from 'normal' research, as reflected by a low $P_{top\ 1\%}/P_{top\ 10\%}$ ratio. In Table 1 (combining four technologies), the USA's advantage over the EU is 2.8-fold when measured using $P_{top\ 1\%}$, but 18.8-fold when measured using $P_{top\ 0.01\%}$.

The failure of the EC to recognize the EU's weaknesses in breakthrough research has economically important consequences because it has prevented the implementation of research policies aimed at addressing these weaknesses. For example, the automotive industry has long been a cornerstone of the EU economy but is now under threat as electric vehicles replace those powered by internal combustion engines (Draghi, 2024b). In this transition, electric batteries—which are also essential for many other applications—play a central role. However, EU breakthrough research in lithium batteries, which was competitive during 2004–2007, has become much less competitive since then (Fig. 7), while the EC proclaimed unsupported successes.

### *5.3. Success depends on efficiency and size*

As already explained, size is an important factor in breakthrough research (Section 1.4). The UK is not a small country and has a highly efficient research system (Fig. 5). It also has the highest number of Nobel Prizes per capita (Gros, 2018). Nevertheless, Table 4 shows that the UK cannot compete with the USA in breakthrough technological research. Even if 100 researchers in the UK had a higher probability than 100 researchers in the USA of developing a new photovoltaic cell or lithium battery, the much larger number of researchers in the USA makes it impossible for the UK to achieve scientific advances at the US rate. To compensate for the difference in size and compete on equal terms in $P_{top\ 0.01\%}$, the UK's $P_{top\ 1\%}/P_{top\ 10\%}$ ratio would have to be substantially higher than that of the USA (Eq. 1). For example, it would need to be approximately 0.33 when the corresponding US value is 0.17, which is unrealistic.

The competition between China and the USA is substantially different because China's $P_{top\ 1\%}/P_{top\ 10\%}$ ratios remain relatively low. In 2025, in the technological topics examined here, China published approximately 13 times more papers than the USA (Fig. 11). If China's $P_{top\ 1\%}/P_{top\ 10\%}$ ratio is 0.09 (Table 1), the USA would need a $P_{top\ 1\%}/P_{top\ 10\%}$ ratio of 0.22 to compensate for China's 13-fold higher publication output. Similarly, if China's ratio were 0.10, the USA would require a $P_{top\ 1\%}/P_{top\ 10\%}$ ratio of 0.24 to offset the same publication advantage. Such high $P_{top\ 1\%}/P_{top\ 10\%}$ ratios were achievable for the USA a few years ago but may become unattainable in the future (Kahn et al., 2026; Science News Staff, 2026).

*5.4. Illusory successes bewilder research policy*

The example above indicates that the only way for the EU to compete with China in high-tech industries is to increase both its size and its research efficiency. The first measure requires better integration of the research systems of the EU member states, as well as the full integration of those of the UK and Switzerland, which would also benefit these two countries. Such greater integration is consistent with the proposals made by Letta (2024) and Draghi (2024a) for the integration of European markets.

A research system integrating the EU, the UK, and Switzerland would be considerably larger than that of the USA. Consequently, the calculations presented in the previous section indicate that Europe's inability to compete effectively with China in high-tech innovation is due primarily to the low efficiency of research in the EU (Tables 1 and 2). Improving breakthrough research in high-tech topics is therefore an urgent priority for the EU. Europe cannot compete with China while its $P_{top\ 1\%}/P_{top\ 10\%}$ ratios remain lower than those of China.

However, achieving such improvement will not be easy. It would require the EC to acknowledge a long-standing weakness in EU research, which would also mean recognizing that its research policy has been misguided for more than 25 years, and to undertake a major reorganization of the EC research services (Jolivet, 2024). Regardless of the origin or accuracy of the well-known statement, *"If you can't measure it, you can't manage it,"* it is evident that research policy cannot be effective if it is based on misleading conclusions, inappropriate indicators, and the neglect of well-founded academic criticism.

Figures 3 and 5 clearly show that the EU and its largest member states—Germany, France, Italy, and Spain—have relatively low research efficiency and that this has not improved over the 14-year period covered by the *Leiden Ranking*. Given that Australia substantially improved the performance of its universities during the past decade (Fig. 6), it is regrettable that Germany, France, Italy, and Spain have not achieved similar progress.

Misleading conclusions about research performance are pervasive in EU reports (Section 1.1). The case of graphene is particularly illustrative because reports on the Graphene Flagship initiative repeatedly refer to European leadership in graphene. For example: “the Graphene Flagship was funded to ensure that Europe would maintain its lead in graphene research and innovation” (https://graphene-flagship.eu/about/first-10-years,

accessed on June, 2026), and “Its core mission is to translate Europe’s scientific leadership in graphene and 2D materials into real-world technologies” https://graphene-flagship.eu/media/sgfbqek5/graphene-flagship-ar-2025-web-spreads-120-ppi-srgb.pdf, accessed on June, 2026). These unsupported claims of leadership are contradicted by the data presented in Table 2 and Fig. 8. Until 2018, leadership was shared by China and the USA; since 2020, China has been the undisputed leader.

This illusion of research success probably exists in many EU member states and may help explain the stagnation of their largest research systems. It is evident in Spain, where, in 2022, the Spanish Parliament unanimously passed a Science Law that effectively enshrines a Spanish version of the European Paradox:

> The Spanish Science, Technology, and Innovation System has achieved research excellence standards perfectly aligned with its economic and geopolitical position on the international stage. However, this excellence in scientific output has not yet been effectively transferred to the productive sector (translated from Spanish; (Jefatura-del-Estado-Español, 2022).

Unfortunately, this illusion is not confined to policymakers but also extends to scientists. A recent manifesto by the Confederation of Spanish Scientific Societies stated:

> The Spanish scientific community, represented by the Confederation of Scientific Societies (COSCE), wishes to express its conviction that the Spanish System of Science, Technology and Innovation (SECTI) has reached standards of research excellence perfectly comparable to those of the most advanced countries (translated from Spanish, https://cosce.org/wp-content/uploads/2026/04/Manifiesto-COSCE-por-un-SECTI-mas-eficiente_rev.pdf, accessed on June 2026).

On the basis of such misconceptions, the EC does not perceive improving research efficiency as a priority. Consequently, although the principal factors that reduce the production of breakthrough research are well known (Besancenot & Vranceaunu, 2024; Rodríguez-Navarro, 2009), the EC has not implemented policies aimed at addressing these problems.

In contrast to the limited attention devoted to improving breakthrough research, increasing R&D investment is often presented as the preferred solution for closing the gap in disruptive innovation. The target of investing 3% of GDP in R&D appears in numerous EC documents (European Commission, 2017). However, before increasing research investment, an important question should be addressed: should the EU continue investing in research aimed at achieving disruptive innovation if its probability of success relative to China is very low? It should be recognized that funding research that produces high-quality 'normal' research but little breakthrough research may ultimately benefit other countries (Section 1.4). For example, what are the benefits for countries investing in research on the next generation of electric batteries if those batteries are ultimately most likely to result from research conducted in China or the USA (Table 1; Fig. 7) and to be commercialized by Chinese or US companies? In high-tech sectors, where innovation depends on breakthrough research that pushes the frontiers of knowledge, continued investment may not be a rational strategy if the probability of success is too low.